\pdfoutput=1
\documentclass[12pt]{article}
\usepackage[colorlinks,linkcolor=blue,citecolor=blue,urlcolor=blue,bookmarks,bookmarksnumbered]{hyperref}
\usepackage{amsmath,amssymb,amsfonts,mathrsfs}
\usepackage[paper=letterpaper,margin=1.0in]{geometry}
\usepackage{graphicx,xcolor}
\RequirePackage{tikz}
 \usetikzlibrary{arrows}
 
\usepackage{cite}

\newcommand{\be}{\begin{equation}}
\newcommand{\ee}{\end{equation}}
\newcommand{\bea}{\begin{eqnarray}}
\newcommand{\eea}{\end{eqnarray}}
\newcommand{\ba}{\begin{array}}
\newcommand{\ea}{\end{array}}
\newcommand{\ben}{\begin{enumerate}}
\newcommand{\een}{\end{enumerate}}
\newcommand{\bi}{\begin{itemize}}
\newcommand{\ei}{\end{itemize}}
\newcommand{\bc}{\begin{center}}
\newcommand{\ec}{\end{center}}
\newcommand{\bfig}{\begin{figure}}
\newcommand{\efig}{\end{figure}}
\newcommand{\bq}{\begin{quotation}}
\newcommand{\eq}{\end{quotation}}
\newcommand{\bt}{\begin{table}}
\newcommand{\et}{\end{table}}
\newcommand{\btab}{\begin{tabular}}
\newcommand{\etab}{\end{tabular}}
\newcommand{\bs}{\begin{slide}}
\newcommand{\es}{\end{slide}}

\let\SS=\S 

\newcommand{\vev}[1]{\langle #1 \rangle}

\newdimen\lft\lft=0pt

\newcommand{\rd}{\mathrm{d}}

\let\tw=\tilde
\let\Tw=\widetilde
\let\a=\alpha

\let\ba=\overline

\let\b=\beta

\let\j=\psi

\let\l=\lambda

\let\p=\pi

\let\vd=\partial
\def\tx{\tilde{x}}

\let\w=\omega

\def\frc#1#2{{\textstyle{#1\over#2}}}

\def\resetby#1#2{\@addtoreset{#2}{#1}}
\def\seceq{\@addtoreset{equation}{section}
              \def\theequation{\thesection.\arabic{equation}}}

\let\sss=\scriptscriptstyle

\def\Dj{\rule[.75ex]{4pt}{.6pt}\kern-4ptD}
\def\dj{d\kern-4pt\rule[1.15ex]{4pt}{.5pt}}
\def\tx{{\mathord{\tilde x}}}

\def\brb#1#2{[\![\mkern2mu#1,\mkern-2mu#2\mkern2mu]\!]}

\makeatletter
\def\section{\@startsection{section}{1}{\z@}
         {-2ex plus-1ex minus-.2ex}{1pt plus1pt}{\large\bfseries\boldmath}}
\def\subsection{\@startsection{subsection}{2}{\z@}
          {-1.5ex plus-1ex minus-.2ex}{0.01pt plus1pt}{\slshape}}
\def\subsubsection{\@startsection{subsubsection}{3}{\z@}
          {-1ex plus-1ex minus-.2ex}{0.01pt plus0.2pt}
          {\small\bfseries\boldmath}}
\def\paragraph{\@startsection{paragraph}{4}{\z@}
          {.75ex \@plus.5ex \@minus.2ex}{-2mm}{\bfseries\boldmath}}
\def\subparagraph{\@startsection{paragraph}{4}{\z@}
          {-.5ex \@plus.33ex \@minus.15ex}{-2mm}
          {$\blacktriangleright\,$\slshape}}
\makeatother

\parskip\medskipamount
\begin{document}

\noindent
\begin{center}
\renewcommand*{\thefootnote}{\fnsymbol{footnote}}
\vspace{7mm}
{\Large\bf Mirror Symmetry, Born Geometry and String Theory}\\
{\large\bf\color[rgb]{.3,.6,.1}--- The Meta Mirror Map ---}
\vspace{3mm}

{{\bf
	Per Berglund${}^{1*}$,
	Tristan H{\"u}bsch${}^{2\dag}$
and
	Djordje Minic${}^{3\ddag}$
}}
\vspace{1mm}

{\footnotesize\it
${}^1$Department of Physics and Astronomy, University of New Hampshire,
 Durham, NH 03824, U.S.A. \\
${}^2$Department of Physics and Astronomy, Howard University, Washington,
 D.C., 20059, U.S.A. \\
${}^3$Department  of Physics, Virginia Tech, Blacksburg, VA 24061, U.S.A.\\
 $^*$\,{\tt per.berglund@unh.edu}, 
 $^\dag$\,{\tt thubsch@howard.edu},
 $^\ddag$\,{\tt dminic@vt.edu}\\
}
\end{center}

\begin{abstract}\noindent
All known string theory models may be obtained as partial fermionization, projection and background Ans{\"a}tze from the original, purely bosonic string theory. The latter theory in turn has been recently shown to describe a chirally and non-commutatively doubled and manifestly T-dual target spacetime. We show herein that this, so-called {\em\/metastring\/} theory automatically includes mirror symmetry. 
\end{abstract}

\paragraph{Introduction:}
A closer look at the underlying structures in string theory shows that the familiar and (almost) independent left- and right-moving degrees of freedom span a chirally doubled phase space-like non-commutative and modular target spacetime $\mathscr{M}$~\cite{Freidel:2015pka,Freidel:2017xsi}. This so called {\em\/Born geometry\/} is specified by (i) the symplectic structure $\omega$, (ii) the bi-orthogonal metric $\eta$, and (iii) the doubly Lorentzian metric $H$. More precisely, this {\em\/metastring\/} target spacetime, $\mathscr{M}$, is an {\em\/almost symplectic and para-Hermitian manifold\/}\footnote{Such spaces admit a signature-$(d,d)$ metric $\eta$ and a symplectic structure $\w$ such that $K\!:=\!\eta{\cdot}\w$ is an {\em\/almost product structure,} $K^2=+{\bf 1}$. This $\eta$ need not be flat, and {\em\/almost\/} means that $\w$ need not be a closed 2-form. Owing to the almost symplectic structure, the existence of the corresponding flat (Bott) connection guarantees that a foliated space is everywhere locally a product of two half-dimensional affine spaces.} that has a compatible {\em\/foliation\/}~\cite{Freidel:2018tkj}: Locally at every point, $p$, $\mathscr{M}$  
is of the form $\mathscr{M}_p=M_p{\times}\Tw{M}_p$, with local coordinates  $(x^a,\tx_b)_p\in M_p{\times}\Tw{M}_p$, where $x^a\!:=\!x^a_{\sss L}{+}x^a_{\sss R}$ and $\tx^a\!:=\!x^a_{\sss L}{-}x^a_{\sss R}$ are combinations of the zero-modes of the corresponding worldsheet scalar fields. For local diffeomorphisms (implemented by the Dorfman generalization of the Lie derivative) to be integrable to finite translations, 
the so-called ``section condition'' is imposed, which halves the spacetime akin to the quantum-mechanical restriction of the classical phase space, $(p,q)$, in the coordinate or momentum representation --- or indeed any other $\p\!:=\!(\a p{+}\b q)$-{\em\/polarization,} as familiar from Geometric Quantization program~\cite{rNH-GQ,rNW-GQ}.

\paragraph{Mirror Doubling:}
A hallmark of the above structure is that
 $\p[T_\mathscr{M}]=(T\oplus T^*)_{\p[\mathscr{M}]}$:
a polarization of (the total space of) the tangent bundle on $\mathscr{M}$ is the generalized/doubled tangent bundle on that polarization of $\mathscr{M}$. 
 In local coordinates of the $\p_x$-polarization, $\p_x[\mathscr{M}]\,{=}\,M_x$, elements of $T_\mathscr{M}$ are given as $v^a(x,\tx)\,\vd_a{+}w_a(x,\tx)\,\rd x^a$, as $\tx$ are locally constant ($\tw\vd^a,\rd\tx_a\!\mapsto\!0$)~\cite{Freidel:2018tkj}. 
 In turn, the same element is $v^a(x,\tx)\,\rd\tx_a{+}w_a(x,\tx)\,\tw\vd^a$ in the $\p_\tx$-polarization, $\p_\tx[\mathscr{M}]\,{=}\,\Tw{M}_\tx$, where $x$ are locally constant and $\vd_a,\rd x^a\!\mapsto\!0$.
 Therefore, swapping the polarization $M_x\leftrightarrow \Tw{M}_\tx$ explicitly identifies $T_{M_x}=T^*_{\Tw{M}_\tx}$ and $T^*_{M_x}=T_{\Tw{M}_\tx}$ --- which is the underlying premise of mirror symmetry. 
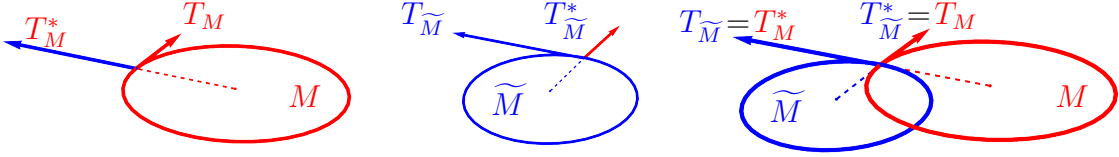
\begin{figure}[htbp]
 \begin{center}
  \begin{picture}(160,15)(-3,5)
   \put(0,-1){\includegraphics[width=50mm]{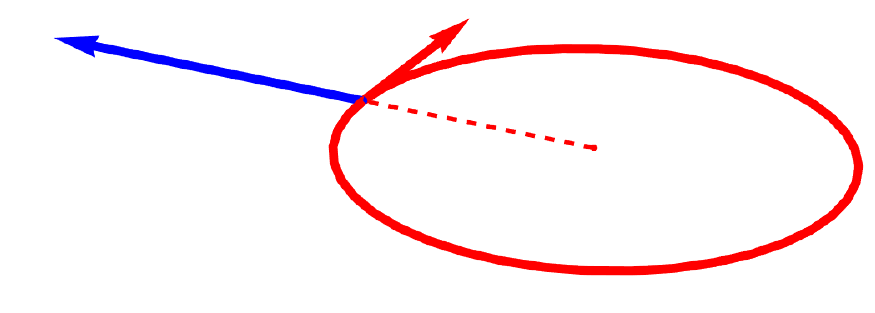}}
    \put(41,6){\color{red}$M$}
    \put(27,17){\color{red}$T_M$}
    \put(6,15){\color{red}$T^*_M$}
   \put(60,0){\includegraphics[width=50mm]{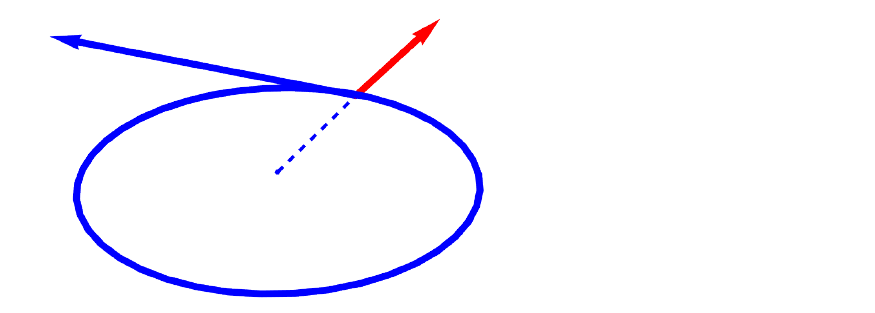}}
    \put(68,5){\color{blue}$\Tw{M}$}
    \put(75,17){\color{blue}$T^*_{\Tw{M}}$}
    \put(56,17){\color{blue}$T_{\Tw{M}}$}
   \put(97,-2){\includegraphics[width=55mm]{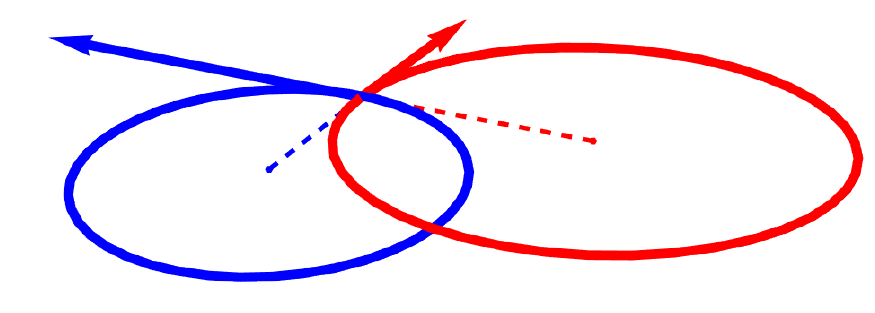}}
    \put(143,6){\color{red}$M$}
    \put(105,4){\color{blue}$\Tw{M}$}
    \put(93,16){${\color{blue}T_{\Tw{M}}}\!=\!{\color{red}T^*_M}$}
    \put(117,17){${\color{blue}T^*_{\Tw{M}}}\!=\!{\color{red}T_M}$}
  \end{picture}
 \end{center}
 \caption{The original $(M,T_M,T^*_M)$, its mirror $(\Tw{M},T_{\Tw{M}},T^*_{\Tw{M}})$, and both depicted together}
 \label{f:M+MM}
\end{figure}
Displacing $M$ and $\Tw{M}$ for clarity, these relationships are sketched in Figure~\ref{f:M+MM}, and corroborate the argument presented in~\cite[\SS\,5.1]{rBHM7}.
 The Born geometry of $\mathscr{M}$  
 explicitly includes mirror symmetry by virtue of having to {\em\/polarize\/} on one half of $\mathscr{M}$, 
 such that polarizing on the complementary half swaps the corresponding tangent and cotangent spaces. This provides a non-commutative version of the {\em\/generalized Calabi-Yau space\/} framework~\cite{Hitchin:2003wq} for the {\em\/entire\/} spacetime, resonating with the recent proposal~\cite{rBHM12}.

\paragraph{Born Heterosis:}
The metastring intrinsic mirror symmetry applies also to an adaptation of the heterotic Ansatz~\cite{Gross:1984dd}, which relies on a partial {\em\/fermionization\/} and supersymmetry: 
On the worldsheet, every chiral boson maps non-locally and nonlinearly but exactly 1--1 to a (chiral) fermion~\cite{Coleman:1974bu,Mandelstam:1975hb}; also~\cite{Witten:1984vr,Harada:1989qp,vonDelft:1998pk,Senechal:1999us}.
 A suitable superconformal action for the resulting collection of worldsheet scalar and fermion fields then induces supersymmetry in the target space, leading to the familiar result that all the various superstring models may be obtained from the bosonic string~\cite{Casher:1985ra}, all of which eliminate the tachyonic instability of the purely bosonic string; see also~\cite{Siegel:1993th}.

Re-bosonizing and adapting the original heterotic Ansatz~\cite{Gross:1984dd,Casher:1985ra}, we partition the 26+26 chiral bosons of the bosonic (meta)string as follows, by identifying $x^0,x^1,\tx^0,\tx^1$ in the light-cone gauge with the worldsheet (chirally doubled) coordinates:
\begin{equation}\mkern-72mu
  \vcenter{\hbox{\begin{tikzpicture}[scale=1.1]
     \path[use as bounding box](-1.2,0)--(12.5,-1.6);
       \foreach\xx in {0,...,9}
        \path(\xx*.5,-.5)node{\footnotesize$x^{\xx}$};
       \foreach\xx in {10,...,17}
        \path(\xx*.5+.1,-.5)node{\footnotesize$x^{\xx}$};
       \foreach\xx in {18,...,25}
        \path(\xx*.5+.2,-.5)node{\footnotesize$x^{\xx}$};
       \foreach\xx in {0,...,9}
        \path(\xx*.5,-1)node{\footnotesize$\tx^{\xx}$};
       \foreach\xx in {10,...,17}
        \path(\xx*.5+.1,-1)node{\footnotesize$\tx^{\xx}$};
       \foreach\xx in {18,...,25}
        \path(\xx*.5+.2,-1)node{\footnotesize$\tx^{\xx}$};
       \draw[green!70!black,densely dashed,thick](-.2,-.3)rectangle++(.9,-.9);
        \path[green!70!black](.3,-.1)node{\scriptsize WS};
       \draw[red,thick](.8,-.3)rectangle++(3.9,-.43);
        \path[red](2.75,-.1)node{\scriptsize light-cone $\perp$-coordinates on $M$};
       \draw[red,thick](4.8,-.3)rectangle++(4.05,-.43);
        \path[red](6.75,-.1)node{\scriptsize light-cone
         $T_M={\color{blue}T^*_{\Tw{M}}}$-fiber};
       \draw[red,thick](8.92,-.3)rectangle++(4.05,-.43);
        \path[red](10.75,-.1)node{\scriptsize Simple pos.\ roots of $E_8$};
       \draw[blue,thick](.8,-.79)rectangle++(3.9,-.43);
        \path[blue](2.75,-1.45)node{\scriptsize light-cone
         $\perp$-coordinates on $\Tw{M}$};
       \draw[blue,thick](4.8,-.79)rectangle++(4.05,-.43);
        \path[blue](6.75,-1.45)node{\scriptsize light-cone
         $T_{\Tw{M}}={\color{red}T^*_M}$-fiber};
       \draw[blue,thick](8.92,-.79)rectangle++(4.05,-.43);
        \path[blue](10.75,-1.45)node{\scriptsize Simple pos.\ roots of $\Tw{E}_8$};
                 \end{tikzpicture}}}
 \label{e:HMS}
\end{equation}
For $a=10,\cdots,17$, the chiral bosons $x^a,\tx^a$ are the preimages of 8+8 fermions, $\j^a,\tw\j^a$, that are induced by supersymmetry to span the indicated (co)tangent bundles. The worldsheet action for these $x^a,\tx^a$ must reflect the re-bosonization of the supersymmetric structure of the action involving $\j^a,\tw\j^a$.
 In turn, for $a=18,\cdots,25$, $x^a,\tx^a$ are compactified on two copies of the
 $\mathbb{R}^8/\Lambda(E_8)$ torus. Their dynamics is the (nonlocal!) re-bosonization of the structure specified for their fermionized counterparts, $x^a\mapsto\l^a$ and $\tx^a\mapsto\tw\l^a$,
adapting the standard description~\cite{Gross:1984dd,Casher:1985ra}. The worldsheet actions are indeed much simpler (and local!) in the {\em\/fermionized picture\/} of $x^a,\tx^a$ for $a>9$; the re-bosonized specification given here however highlights the implications of the metastring Born geometry.

The {\em\/non-commutative\/} metastring structure of $\mathscr{M}$  
then has the following immediate implications:
\begin{enumerate}\vspace{-5pt}\itemsep=-1pt\vspace*{-1mm}

 \item\label{i:1}
  The metastring metric, $\eta_{AB}|_{10\cdots17}$, reproduces the canonical fiber-wise pairing of the two standard bundles:
  $\vev{T_M,T^*_M}{=}\vev{\smash{T^*_{\sss\Tw{M}}},\smash{T_{\sss\Tw{M}}}}{=}1$.
  
 \item\label{i:2}
  The metastring metric, $\eta_{AB}|_{18\cdots25}$, implies a dual pairing between the $E_8$'s simple positive roots and those of $\Tw{E}_8$. This implies that the roots of $\Tw{E}_8$ are $\eta$-canonically reciprocal to those of $E_8$ --- which identifies $\Tw{E}_8$ as the {\em\/Langlands dual,} $E^\vee_8$~\cite{Borel:1979ux}, i.e., the electro-magnetic dual~\cite{Kapustin:2006pk,Frenkel:2009ra}! Being simply laced, $E_8^\vee\approx E_8$, and they are physically indistinguishable. The $E_8$-group elements are exponential functionals of the $\{x^{18},\cdots,x^{25}\}$, which can be arranged to commute\cite{Freidel:2015pka,Freidel:2017xsi} with the $\Tw{E}_8$-group exponential functionals of the $\{\tx^{18},\cdots,\tx^{25}\}$, realizing the $E_8\!\times\!\Tw{E}_8$ Yang-Mills gauge group as expected.

 \item\label{i:3}
For $a=2,\,{\cdots}\,,9$, the diagram~\eqref{e:HMS} identifies
 $x^a,\tx^a$ with the light-cone {\em\/base\/} coordinates,
 $x^{a+8}\!\mapsto\!\rd x^a\!=\!\tw\vd^a$ and $\tx^{a+8}\!\mapsto\!\vd_a\!=\!\rd\tx_a$
 with the light-cone {\em\/fibre\/} coordinates (as befits the preimages of fermions typically spanning the tangent and cotangent bundles), and
 $(x^{a+16},\tx^{a+16})\mapsto(\rho_a,\rho_{\tw a})$
 with the simple positive roots.
Then:
\begin{enumerate}\itemsep=-1pt\vspace*{-.25\baselineskip}

 \item
  All target spacetime fields are a priori bi-local: they {\em\/depend\/} on both
  $x^\mu$ and $\tx_\mu$.

 \item They are {\em\/valued\/} as follows:
  $A_\mu\!^\a$ in $T^*_M{\times}E_8$ and
  $A^{\mu\tw\a}$ in $T_M{\times}\Tw{E}_8$, the latter of which define
  $A_\mu\!^{\tw\a}{:=}H_{\mu\nu}A^{\nu\tw\a}$ in $T^{\,\flat}_{\!M}{\times}\Tw{E}_8$ using the metastring $H_{AB}$-metric.
 \vspace*{-.5\baselineskip}
\end{enumerate}
 These originally bi-local fields then give rise to both an $E_8$-gauge field {\em\/plus\/} the dual (``dark'') field valued in the {\em\/same\/} $E_8$ algebra. With the analogous for the $\Tw{E}_8$-valued gauge fields, we have both $A_\mu{}^\a(x),A_\mu{}^{\tw\a}(x)$ and  their duals, $\Tw{A}_\mu{}^\a(\tx),\Tw{A}_\mu{}^{\tw\a}(\tx)$.

 \item\label{i:4}
  The metastring non-commutativity implies the correlation,
  $\frc{\l_4\!^2}{L^4}\int_x\mathfrak{F}^{\mu\nu}\brb{A_\mu^\a(x)}{\Tw{A}_\nu^{\,\tw\b}(\tx)}\eta_{\a\tw\b}$, between the {\em\/visible\/} $E_8$-fields and the {\em\/dark\/} $\Tw{E}_8$-fields, as well as the analogous correlation term with $E_8\leftrightarrow\Tw{E}_8$ swapped~\cite{rBHM10}.

 \vspace*{-.5\baselineskip}
\end{enumerate}
 The above statements in~\ref{i:3} insure that our Ansatz~\eqref{e:HMS} reproduces the standard gauge field content of the heterotic string. The statements~\ref{i:1} and~\ref{i:2} however show that the Born geometry is not only perfectly aligned with the standard geometry and dynamics in the heterotic string theory, but additionally links the two copies of $E_8$ as each other's Langlands dual. Finally the correlation~\ref{i:4} is a straightforward but novel result~\cite{rBHM10}.

\paragraph{Outlook:}
In conclusion, we comment on various ramifications of
this new view of mirror symmetry. Our discussion of mirror symmetry 
and its relation to  Born geometry of an intrinsically non-commutative 
and T-duality covariant formulation of string theory can be naturally 
related  to the old observation that T-duality is deeply related to mirror symmetry~\cite{rSYZ-Mirr}. In that case, as in our discussion, one should pay close attention to the issues of the local versus global formulation of mirror symmetry as T-duality. In particular, our phase space treatment of string theory, and mirror symmetry, as encapsulated in Born geometry, naturally relates to homological mirror symmetry involving the (derived) Fukaya categories (symplectic structure) on one side and the (derived) category of coherent sheaves (complex structure)~\cite{Kontsevich:1995wk}.
Also, the world-sheet formulation of metastring theory should be 
related to the topological A and B model projections~\cite{Witten:1991zz}, 
so that these appear as gauge fixed versions 
of the metastring formulation that explicitly incorporates Born geometry.
Finally, we would like to understand non-perturbative aspects of this
approach in the context of a purely bosonic but non-commutative 
(and, in principle, non-associative) matrix-model like formulation; see~\cite{rBHM10}.

\paragraph{Acknowledgments:}
 We would like to thank the organizers of the 2021 Nankai Chern symposium, and Yang-Hui He in particular for this extraordinary opportunity.
 DM thanks Laurent Freidel and Rob Leigh for numerous insightful discussions over many years on the topic of quantum gravity and string theory.
 PB would like to thank the CERN Theory Group for their hospitality over the past several years.
 TH is grateful to the Department of Physics, University of Maryland, College Park MD, and the Physics Department of the Faculty of Natural Sciences of the University of Novi Sad, Serbia, for the recurring hospitality and resources.
 DM is grateful to Perimeter Institute for hospitality and support.
The work of PB is supported in part by the Department of Energy 
grant DE-SC0020220.
 The work of DM is supported in part by Department of Energy 
(under DOE grant number DE-SC0020262) and the Julian Schwinger Foundation.

\begingroup
\footnotesize\baselineskip=13pt \parskip=0pt plus2pt minus1pt

\endgroup

\end{document}